\documentclass{aa}

\usepackage{graphicx}
\usepackage{txfonts}
\begin{document}

\title{External reconnection and resultant reconfiguration of overlying magnetic fields during sympathetic eruptions
of two filaments}

\author{Y. J. Hou \inst{1,2}
      \and T. Li \inst{1,2}
      \and Z. P. Song \inst{3,1}
      \and J. Zhang \inst{3,1}
      }

\institute{CAS Key Laboratory of Solar Activity, National Astronomical Observatories,
           Chinese Academy of Sciences, Beijing 100101, China; yijunhou@nao.cas.cn\\
           \and School of Astronomy and Space Science, University of Chinese Academy of Sciences, Beijing 100049, China\\
           \and School of Physics and Materials Science, Anhui University, Hefei 230601, China; zjun@ahu.edu.cn
           }

\date{Received ????; accepted ????}


  \abstract
   {Sympathetic eruptions of two solar filaments have been studied for several decades, but the detailed physical
   process through which one erupting filament triggers another is still under debate.
   }
   {Here we aim to investigate the physical nature of a sympathetic event involving successive eruptions of two
   filaments on 2015 November 15-16, which presented abundant sympathetic characteristics.
   }
   {Combining data from the \emph{Solar Dynamics Observatory} and other observatories as well as results of nonlinear
   force-free field (NLFFF) extrapolations, we study the evolution of observational features and magnetic topology
   during the sympathetic event.
   }
   {The two filaments (north F1 and south F2) were separated by a narrow region of negative polarity, and F1 firstly
   erupted, producing a two-ribbon flare. When the outward-spreading ribbon produced by F1 approached stable F2, a
   weak brightening was observed to the south of F2 and then spread northward, inward approaching F2. Behind this
   inward-spreading brightening, a dimming region characterized by a plasma density reduction of $30\%$ was extending.
   NLFFF extrapolations with a time sequence reveal that fields above pre-eruption F1 and F2 constituted a quadrupolar
   magnetic system with a possible null point. Moreover, the null point kept moving towards F2 and descending within the
   following hours. We infer that the rising F1 pushed its overlying fields towards the fields above stable F2 and caused
   successive external reconnection between the overlying fields. From outside to inside (lower and lower in height), the
   fields above pre-eruption F2 were gradually involved in the reconnection, manifesting as the inward-spreading
   brightening and extending dimming on the south side of F2. Furthermore, the external reconnection could reconfigure
   the overlying fields of F2 by transporting magnetic flux from its west part to the east part, which is further
   verified by the subsequent partial eruption of F2.
   }
   {We propose an integrated evidence chain to demonstrate the critical roles of external magnetic reconnection and the
   resultant reconfiguration of overlying fields on the sympathetic eruptions of two filaments.
   }


\titlerunning{Sympathetic eruptions, filaments, and external reconnections}
\authorrunning{Hou et al.}

\maketitle
%

\section{Introduction}
Solar filaments are large magnetic structures confining cool and dense plasma suspended in the hot and tenuous corona.
They are seen in absorption in H$\alpha$ and extreme ultraviolet (EUV) channels when observed on the solar disk, and 
appear in emission and are described as ``prominences'' when observed above the solar limb (Mackay et al. 2010; Parenti 
2014). A stable filament could erupt due to the loss of the balance of forces acting on them under several possible
mechanisms, such as magnetic flux emergence and cancellation (Chen \& Shibata 2000; Zhang et al. 2001; Su et al. 2011;
Dacie et al. 2018), tether cutting reconnection (Moore et al. 2001; Chifor et al. 2007; Chen et al. 2014), breakout
reconnection (Sterling \& Moore 2004; DeVore \& Antiochos 2008; Li et al. 2018), and ideal magnetohydrodynamic instability
(Williams et al. 2005; T{\"o}r{\"o}k \& Kliem 2005; Hou et al. 2018; Zou et al. 2020). Filament eruptions are widely
believed to play a critical role in the onset of solar flares and coronal mass ejections (CMEs) (Lin \& Forbes 2000;
Chen 2011; Schmieder et al. 2013).

In two-dimensional (2D) eruption scenarios developed from the classical ``CSHKP'' model (Carmichael 1964; Sturrock 1966;
Hirayama 1974; Kopp \& Pneuman 1976), filament eruptions, solar flares, and CMEs are different observational
manifestations of a more general single eruption process. It is well established that an erupting filament, usually
lying above the magnetic polarity inversion line (PIL), stretches its overlying magnetic fields upwards and form a 
current sheet below, where magnetic reconnection occurs. If escaping successfully, this erupting filament would
produce a CME into the interplanetary space. The particles accelerated by energy released through the reconnection
propagate downward along the newly-formed magnetic field and hit the lower solar atmosphere, producing two bright flare
ribbons on both sides of the PIL (Priest \& Forbes 2002; Shibata \& Magara 2011; Hou et al. 2016). With the progress 
of the reconnection, new groups of field lines successively form with their altitudes increasing and brightened
footpoints (the flare ribbons) apparently separating from each other. Therefore, the morphology and evolution of flare
ribbons or brightenings in \textbf{the} lower solar atmosphere can provide significant insights into topology and
reconfiguration of magnetic system during the solar eruption process (Savcheva et al. 2015; Qiu et al. 2017; Li et al.
2017b). Moreover, coronal dimmings (or transient coronal holes), regions characterized by abrupt emission reduction in
EUV and soft X-ray (SXR) channels, are another distinct phenomenon associated closely with these solar eruptions (Hudson
et al. 1996; Sterling \& Hudson 1997; Zarro et al. 1999; Harrison et al. 2003; Jiang et al. 2011; Zhang et al. 2017;
Veronig et al. 2019). In a recent series of works, two different types of coronal dimmings are distinguished: core and
secondary dimmings (Mandrini et al. 2007; Dissauer et al. 2018a, b). The core dimmings are localized regions with an
impulsive density decrease up to 50$\%$--70$\%$ within half an hour and do not replenish for more than 10 hours, whereas
the secondary dimmings are more shallow and widespread region, which evolve more gradually and starts to recover after
1--2 hours (Vanninathan et al.2018). The core dimmings are interpreted as the footpoints of erupting magnetic flux ropes,
while the secondary dimmings are believed to be caused by the stretch and partial reconnection of overlying fields
(Attrill et al. 2007; Mandrini et al. 2007; Dissauer et al. 2018b).

Besides the single eruption mentioned above, the Sun also ubiquitously breeds sympathetic eruptions, which are defined
as causally-linked eruptions occurring with a relatively short time interval in different but physically related source
regions (Wang et al. 2001; Moon et al. 2002; Wheatland \& Craig 2006; Schrijver \& Title 2011; Titov et al. 2012).
Observational and numerical works suggest that the physical linkages between the sympathetic eruptions should essentially
be of a magnetic nature (Ding et al. 2006; T{\"o}r{\"o}k et al. 2011; Wang et al. 2016). For example, sympathetic
eruptions of two filaments are believed to be caused by the reconnection-related changes of background magnetic fields.
Due to impacts from a nearby erupting filament, constraint fields above the other stable filament would be sufficiently 
removed (Shen et al. 2012; Lynch \& Edmondson 2013; Joshi et al. 2016; Wang et al. 2018). The reconnection between the 
overlying fields applied in this scenario is virtually the same as the concept of external magnetic reconnection in the 
break-out model (Antiochos et al. 1999; Sterling \& Moore 2004; Karpen et al. 2012; Zhou et al. 2017, 2019). 
The external reconnection occurs above the erupting field (e.g., magnetic flux rope) instead of below it as
depicted by the classical 2D ``CSHKP'' model, thus produces various distinctly different observational features. In a
broader meaning, besides between two sets of overlying fields, the external reconnection can also refer to the magnetic
reconnection occurring between one erupting magnetic flux rope and its overlying arcades (Li et al. 2018; Yang et al.
2019b; L{\"o}rin{\v{c}}{\'\i}k et al. 2019; Chen et al. 2019). Such an external reconnection geometry is consistent with
the configuration of so-called three-dimensional (3D) ``ar-rf'' reconnection, which occurs between the leg of an eruptive
flux rope and neighboring inclined arcades (Aulanier \& Dud{\'\i}k 2019).

To verify the sympathetic eruption scenario requiring the occurrence of external reconnection, it is necessary to
find evidence from three aspects as follows: pre-existing magnetic topology in favor of occurrence of the reconnection,
observational features of the reconnection when it happens, and the resultant reconfiguration of background fields.
Although previous works reported some supportive clues such as null point between two filaments involved in sympathetic
eruptions (Zuccarello et al. 2009; Li et al. 2017a), pre-eruption coronal dimming appearing on both sides of the filament
(Jiang et al. 2011; Wang et al. 2018), and bright ribbons detected around the stable filament (Joshi et al. 2016; Wang et
al. 2018), a detailed analysis with an integrated evidence chain for the sympathetic eruptions of two filaments has rarely
been proposed. Aiming to find the causal link between sympathetic eruptions of two filaments, Song et al. (2020) reported
a sympathetic event involving successive eruptions of two filaments (F1 and F2). They provided a qualitative
description about the causal link between the sympathetic eruptions of F1 and F2: the southwest ribbon formed from eruptive
F1 invaded F2 from its southeast region with relatively weaker overlying magnetic fields relative to its northwest region,
disturbing F2 and leading F2 to erupt eventually. However, due to the lack of observational evidence for the interaction
between erupting F1 and pre-eruption F2 as well as the magnetic topology change during the sympathetic eruptions,
some questions still remained to be answered, such as ``How did the southwest ribbon formed by eruptive F1 `invade'
pre-eruption F2?'' and ``How did this intruding flare ribbon disturb F2 and then lead F2 to the eventual eruption?'' To
obtain more insight into the specific physical process of the interaction between erupting F1 and pre-eruption F2, and the
decisive factor of the subsequent eruption of F2, further analysis of more observational evidence and evolution of magnetic
topology needs to be performed in a second study.

In the present work, we investigate the same event as reported by Song et al. (2020) and present integrated evidence for
the sympathetic process from the aspects of magnetic topology before reconnection, signatures during reconnection, and
results after reconnection. The data from the \emph{Solar Dynamics Observatory} (\emph{SDO}; Pesnell et al. 2012) and other
observatories are analyzed to reveal kinematic evolutions of the filaments and various observational signatures formed by
the reconnection, as well as the reconfiguration of overlying fields during the sympathetic eruptions. Based on more
thorough nonlinear force-free field (NLFFF) extrapolations with a time sequence, we reveal the magnetic topology above the
two filaments and its evolution that lead to various features during the eruptions. A more complete and concrete physical
scenario is depicted in the present work to restore the realistic process of the interaction between the two filaments and
the subsequent eruption of F2.

The remainder of this paper is structured as follows. Section 2 describes the observations and methods of data analysis
taken in our study. In Sect. 3, the results of observations and analysis are presented and discussed. Finally, we summarize
the major findings in Sect. 4.

\section{Observations and data analysis}
The sympathetic eruptions of the two filaments (F1 and F2) occurred on 2015 November 15-16, which were well observed by
the \emph{SDO}. The Atmospheric Imaging Assembly (AIA; Lemen et al. 2012) on board the \emph{SDO} provides successive
observations of the multilayered solar atmosphere in 10 channels, including 7 EUV passbands with a cadence of 12 s and
a spatial resolution of 1.{\arcsec}5. The \emph{SDO}/Helioseismic and Magnetic Imager (HMI; Schou et al. 2012) offers
one-arcsecond resolution full-disk intensitygrams, line-of-sight (LOS) magnetograms, and Dopplergrams every 45 s, and
photospheric vector magnetograms at a cadence of 720 s. Here we mainly analyze the AIA 304 {\AA}, 171 {\AA}, 193 {\AA},
211 {\AA}, 335 {\AA}, 94 {\AA}, 131 {\AA} images, HMI LOS magnetograms, and photospheric vector magnetograms. Moreover,
H$\alpha$ observations from the Global Oscillation Network Group (GONG; Harvey et al. 1996), solar corona images taken
by the Large Angle and Spectrometric Coronagraph (LASCO; Brueckner et al. 1995) on board the \emph{SOHO}, and SXR 1--8
{\AA} flux from the \emph{GOES} are also employed.

Based on the almost simultaneous observations of 6 AIA EUV channels (131 {\AA}, 94 {\AA}, 335 {\AA}, 211 {\AA}, 193 {\AA},
and 171 {\AA}), we perform the differential emission measure (DEM) analysis based on the code ``\verb#xrt_dem_iterative2.pro#''
in the Solar Software package (Weber et al. 2004). After modified slightly, this method has been proved to work well with
AIA data for retrieving the plasma DEM (Cheng et at. 2012). The total emission measure (EM) in temperature range
$T_{min} \leq lgT \leq T_{max}$ then can be calculated as follows:
\begin{equation}
EM=\int_{T_{min}}^{T_{max}} DEM(T) dT.
\label{eq1}
\end{equation}
The density $n$ in the structure of interest is evaluated as:
\begin{equation}
n=\sqrt{EM / l},
\label{eq2}
\end{equation}
where $l$ is the LOS depth of the emission structure. Noting that, as investigated in Su et al. (2018), the DEM solutions
derived from the routine ``\verb#xrt_dem_iterative2.pro#'' as well as the other two popular codes: the regularized
inversion code (Hannah \& Kontar 2012) and the sparse inversion code (Cheung et al. 2015), cannot effectively constrain
plasma DEM at flaring temperatures (i.e., showing a significant amount of plasma above 10 MK) with AIA data alone. But in
the present work, what we want to study through the DEM analysis is the coronal dimming region, which is mostly formed by
the evacuation of plasma with a quiet-Sun coronal temperature of several MK (Dissauer et al. 2018a). Considering that this
temperature value is far below the threshold value of 10 MK, we believe the routine modified from
``\verb#xrt_dem_iterative2.pro#'' could be safely applied to AIA data here for a rough estimate of DEM-derived parameters
(e.g., density and temperature).

To understand the 3D magnetic fields above F1 and F2 before their eruptions, we utilize the ``weighted optimization" method
to perform NLFFF extrapolations (Wiegelmann 2004; Wiegelmann et al. 2012) based on the HMI photospheric vector magnetograms
observed at 22:00 UT on November 15 and 00:00 UT on November 16. The NLFFF calculations are performed within a box of
$600\times552\times512$ uniformly spaced grid points ($436\times401\times372$ Mm$^{3}$). Furthermore, we calculate the
distribution of decay index, defined here as:
\begin{equation}
n(z)=-z \frac{dln(B_{hp})}{dz},
\label{eq1}
\end{equation}
in which $B_{hp}$ is the strength of horizontal component of the potential field derived from potential-field source-surface
model (PFSS; Schatten et al. 1969), and $z$ denotes the radial height above the photosphere. Kliem \& T{\"o}r{\"o}k (2006)
has demonstrated that in an analytical model of torus instability, a toroidal current ring becomes unstable when the decay
index of the external poloidal field ($B_{ep}$) reaches a critical value of $\sim$1.5. However, due to the difficulty in
isolating $B_{ep}$, the horizontal component of the potential field ($B_{hp}$) is usually adopted as an approximation since
$B_{hp}$ is orthogonal to the axis of the toroidal current ring and thus can create downward $J \times B$ force (D{\'e}moulin
\& Aulanier 2010; Kliem et al. 2013; Wang et al. 2018).

\section{Results and Discussion}
\subsection{Overview of Sympathetic Eruptions of the Two Filaments}
\begin{figure*}
\centering
\includegraphics [width=0.88\textwidth]{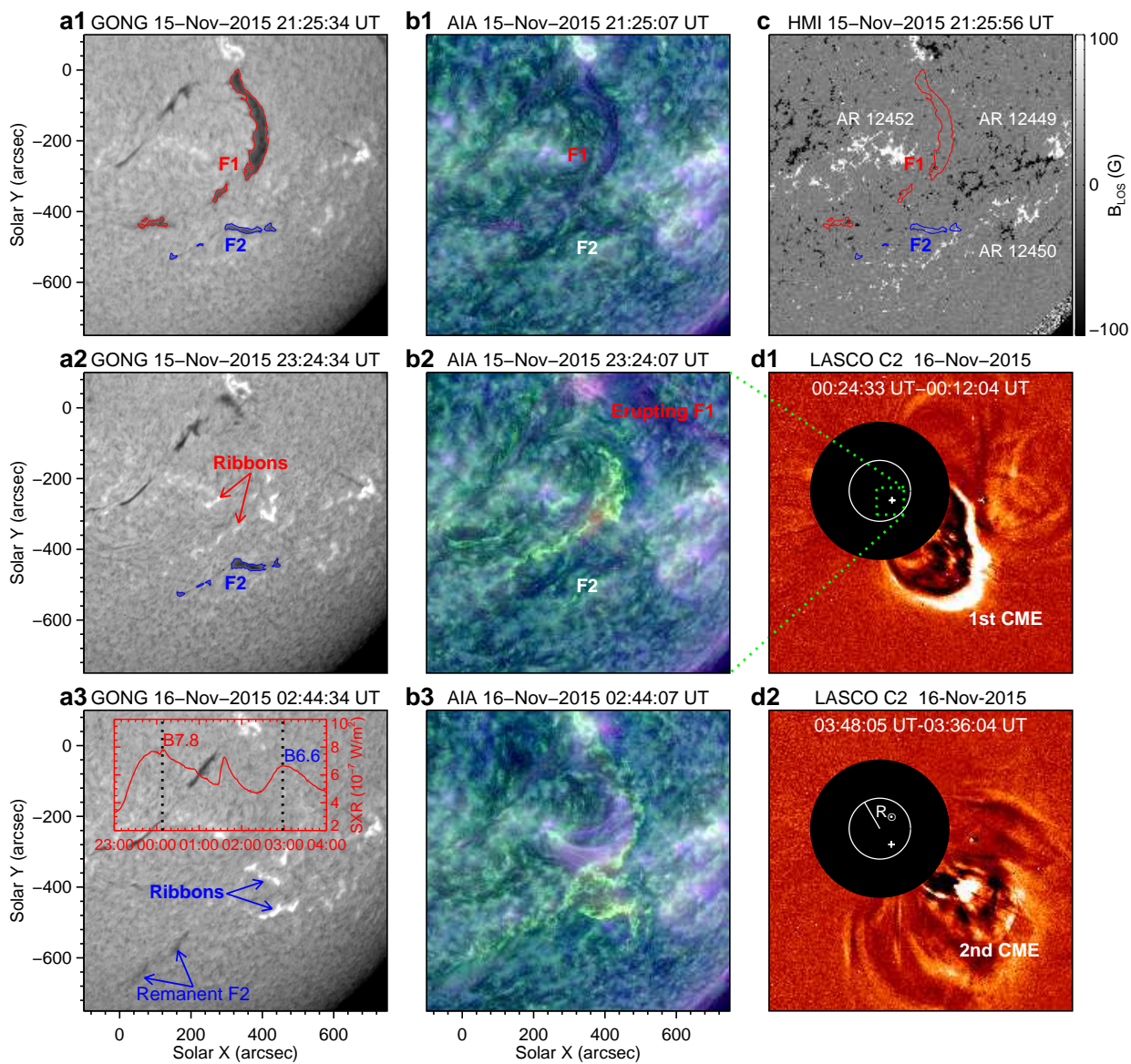}
\caption{Overview of the sympathetic eruptions of two filaments on 2015 November 15-16.
(a1)--(a3): Sequence of GONG H$\alpha$ images displaying evolutions of the two filaments (F1 and F2). The red and
blue contours in panels (a1)--(a2) outline the filament fragments of F1 and F2, respectively. Ribbons produced by
eruptive F1 and F2 and remanent F2 are denoted by arrows in panels (a2)--(a3). The \emph{GOES} SXR 1--8 {\AA} flux
variation overlaid in panel (a3) reveals two B-class flares caused by eruptions of the two filaments.
(b1)--(b3): Corresponding composite images of \emph{SDO}/AIA 131 {\AA}, 304 {\AA}, and 171 {\AA} channels.
(c): Corresponding \emph{SDO}/HMI LOS magnetogram.
(d1)--(d2): Difference images from \emph{SOHO}/LASCO C2 showing two successive CMEs driven by the sympathetic
eruptions of F1 and F2.
An animation (figure1.mov) of H$\alpha$, AIA composite, and LASCO C2 images, covering 21:30 UT on November 15
to 03:54 UT on November 16, is available in the on-line journal.}
\label{fig1}
\end{figure*}

On 2015 November 15, F1 and F2 were located in the southwest quadrant of the solar disk, separated by a narrow
region of negative polarity (top panels of Fig. 1). Figure 1(c) shows that F1 lay above a polarity inversion line (PIL)
region between active regions (ARs) 12452 and 12449, and F2 lay above a PIL within AR 12449. Around 21:30 UT, F1 began
to rise, then accelerated, and eventually erupted southwestward in the plane of the sky (POS). The eruptive F1 produced
a B7.8 two-ribbon flare (panels (a2) and (b2)) and a subsequent CME with a mean projected speed of 501 km s$^{-1}$ and
a width angle of 122{\degr} in the \emph{SOHO} LASCO C2 coronagraph field of view (panel (d1)). At about 00:30 UT on
November 16, F2 started to erupt partially and caused another B6.6 two-ribbon flare and a CME with a mean projected
speed of 853 km s$^{-1}$ and a width angle of 166{\degr} (bottom panels) \footnote{Please see more information about
the CMEs related to the reported event in the \emph{SOHO} LASCO CME catalog (Yashiro et al. 2004), \url{https://cdaw.gsfc.nasa.gov/CME_list/}}. The \emph{GOES} SXR 1--8 {\AA} flux variation in panel (a3) reveals that
the two B-class flares caused by the successive eruptions of F1 and F2 peaked at 00:08 UT and 02:58 UT on November 16,
respectively (also see the associated animation). The peak between them was caused by a flare occurring in AR 12454,
far away from the region of interest.

Based on the closeness in both space and time of the two erupting filament, we can interpret them as sympathetic eruptions.
Studying the same event, Song et al. (2020) presented a detailed analysis of the F1 eruption process and concluded that
it is caused by the flux cancellation within F1 channel (similar to the events reported by Zhang et al. 2001; Hong et al. 
2011; Yang \& Chen 2019; Yang et al. 2019a). They claimed that the non-uniformity of the magnetic fields above F2 could 
be responsible for the sympathetic eruption of F2. In the present work, focusing on more important observational evidence 
(e.g., inward-spreading brightening and dimming region on the south side of the pre-eruption F2, subsequent partial 
eruption of F2, and the bending structure formed between the erupting west part and the remained east part of F2),
we aim to reveal specific physical process of the interaction between erupting F1 and pre-eruption F2 and the subsequent
eruption of F2. The results in this work would significantly supplement the former works about the sympathetic eruptions
of filaments.

\subsection{Inward-spreading Brightening and Coronal Dimming on the South Side of Pre-eruption F2}
\begin{figure*}
\centering
\includegraphics [width=0.99\textwidth]{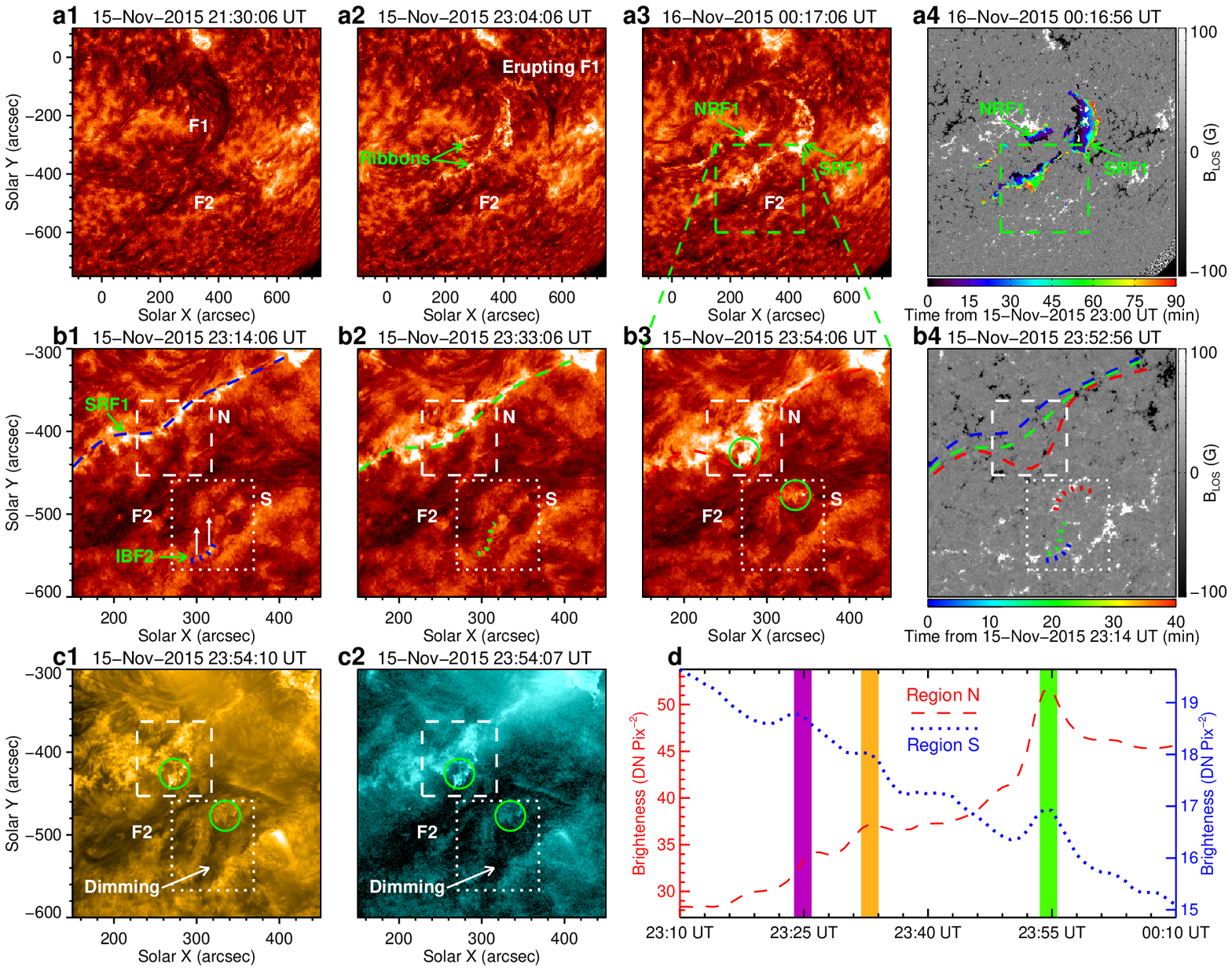}
\caption{Kinematic evolutions of eruptive F1 and the inward-spreading brightening on the south side of pre-eruption F2.
(a1)--(a3): Sequence of AIA 304 {\AA} images showing evolutions of F1 and two flare ribbons (NRF1 and SRF1).
(a4): HMI LOS magnetogram superimposed with colored regions indicating positions of the evolving flare ribbons detected
at different times in 304 {\AA} channel.
(b1)--(b3): Enlarged 304 {\AA} images displaying evolutions of SRF1 (dashed curves) and northward bright ribbon (IRF2)
appearing on the south side of F2 (dotted curves). Green circles in (b3) mark simultaneous brightenings
on opposite sides of F2.
(b4): Similar to (a4), but for the ribbons approaching F2 from two sides.
(c1)--(c2): 171 {\AA} and 131 {\AA} images corresponding to (b3).
(d): Variations of mean emission strength of 304 {\AA} passband within the regions N and S in (b1)--(b3).
An associated animation (figure2.mov) of 304 {\AA}, 171 {\AA}, and 131 {\AA} images, covering from 21:30 UT on
November 15 to 00:30 UT on November 16, is available online.
}
\label{fig2}
\end{figure*}

Sequence of 304 {\AA} images in Figs. 2(a1)-2(a3) display the evolutions of eruptive F1 and the induced two flare
ribbons. To intuitively exhibit apparent motions of the two ribbons, we indicate the positions of newly brightened
ribbons with different colors according to the time of their appearances and then superimpose them on one HMI LOS
magnetogram (panel (a4)). The ribbons on both sides of the PIL related to F1 showed apparent separation motions, i.e.,
the northeast ribbon (NRF1) moved northeastward, and the southwest ribbon (SRF1) spread southwestward. It is pretty
remarkable that in the region outlined by the green square in panel (a3), when SRF1 approached F2, a relatively weak
brightening (IBF2) was observed on the south side of F2 around 23:14 UT on November 15 and then kept spreading northward,
inward approaching F2 (see panels (b1)-(b4) and associated animation). Panel (b4) indicates that from south to north,
this inward-spreading brightening IBF2 swept regions with positive polarity on the south side of F2 and eventually
stopped at a ribbon-shaped positive field close to F2. Behind IBF2, a distinct coronal dimming region was extending
(panels (c1) and (c2)). Then around 23:54 UT, a bright cusp structure ahead of SRF1 and brightenings simultaneously
appeared on opposite sides of F2 (green circles in panels (b3), (c1), and (c2)). Calculating the mean emission strength
of 304 {\AA} channel within regions N and S defined in panel (b1), we notice that the emission flux variations of the
two regions match well with each other in the aspect of transient emission strengthening (colored regions in panel (d)).
The synchronicity feature implies that the brightenings on opposite sides of pre-eruption F2 could result from
the same physical process.

\begin{figure*}
\centering
\includegraphics [width=0.85\textwidth]{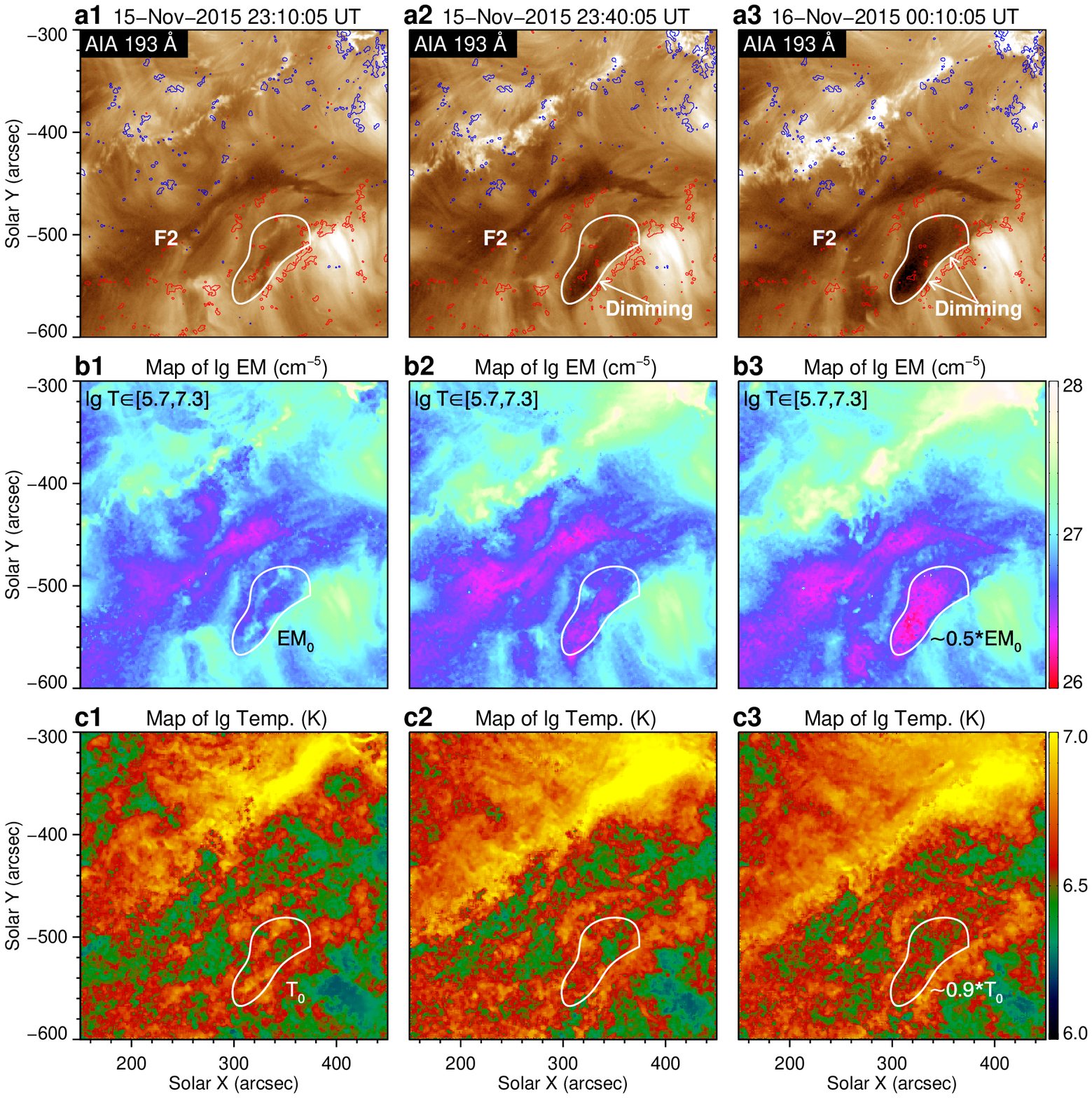}
\caption{Coronal dimming on the south side of pre-eruption F2.
(a1)--(a3): Sequence of AIA 193 {\AA} images showing the appearance of coronal dimming on the south side of F2 before
its eruption. The red and blue curves are contours of the HMI LOS magnetograms at $\pm$50 G. The white curves
outline the border of the dimming region around 00:10 UT on November 16.
(b1)--(c3): Corresponding total EM and DEM-weighted average temperature maps exhibiting the evolution of plasma density
and temperature in the dimming region.
An associated animation (figure3.mov) of 193 {\AA} images, covering from 23:10 UT on November 15 to 00:10 UT on November
16, is available online.
}
\label{fig3}
\end{figure*}

As shown in Fig. 2(d), the emission flux of region ``S'' also present an obvious decline after 23:10 UT, which could
attribute to the coronal dimming occurring on the south side of pre-eruption F2. Here we perform DEM analysis for the
coronal dimming region based on observations of 6 AIA EUV channels and show the results of plasma density and temperature
in Fig. 3. The dimming region outlined by the white curves in Fig.3 was located above the region with positive polarity 
and underwent a noticeable density depletion after 23:10 UT on November 15 (panels (b1)-(b3)). Taking the mean value of 
plasma total EM in the dimming region as $EM_{0}$ around 23:10 UT on November 15, then we measure that the value 
dramatically decreased to $\sim0.5 EM_{0}$ within 1 hour. Here we also employ the sparse inversion code (Cheung et al. 2015;
Su et al. 2018) and find that this method gives similar results. According to Eq. \ref{eq2}, assuming the LOS depth $l$ as
a constant during the evolution of the dimming region, we can derive that the corresponding plasma density would decrease
from $n(t_{0})$ to $n(t) = \sqrt{EM(t)/EM(t_{0})} \times n(t_{0}) \simeq 0.7n(t_{0})$. That is a density reduction of about
$30\%$ within 1 hour. Noting that in the actual situation, the LOS depth $l$ could get larger due to the stretch of magnetic 
fields and the expansion of the plasma in the dimming regions (Veronig et al. 2019). Thus, the value of $30\%$ for the 
density reduction ought to be a lower limit. Additionally, panels (c1)-(c3) denote that the mean value of DEM-weighted
average temperature in the dimming region changed from $T_{0}$ to $\sim0.9 T_{0}$, only slightly decreasing by $10\%$.
These results support the viewpoint that the coronal dimmings are primarily caused by mass expansion (or evacuation) and 
subsequent density depletion rather than temperature decrease (Hudson et al. 1996; Harra \& Sterling 2001; Jin et al. 2009;
Tian et al. 2012; Veronig et al. 2019).

\begin{figure*}
\centering
\includegraphics [width=0.95\textwidth]{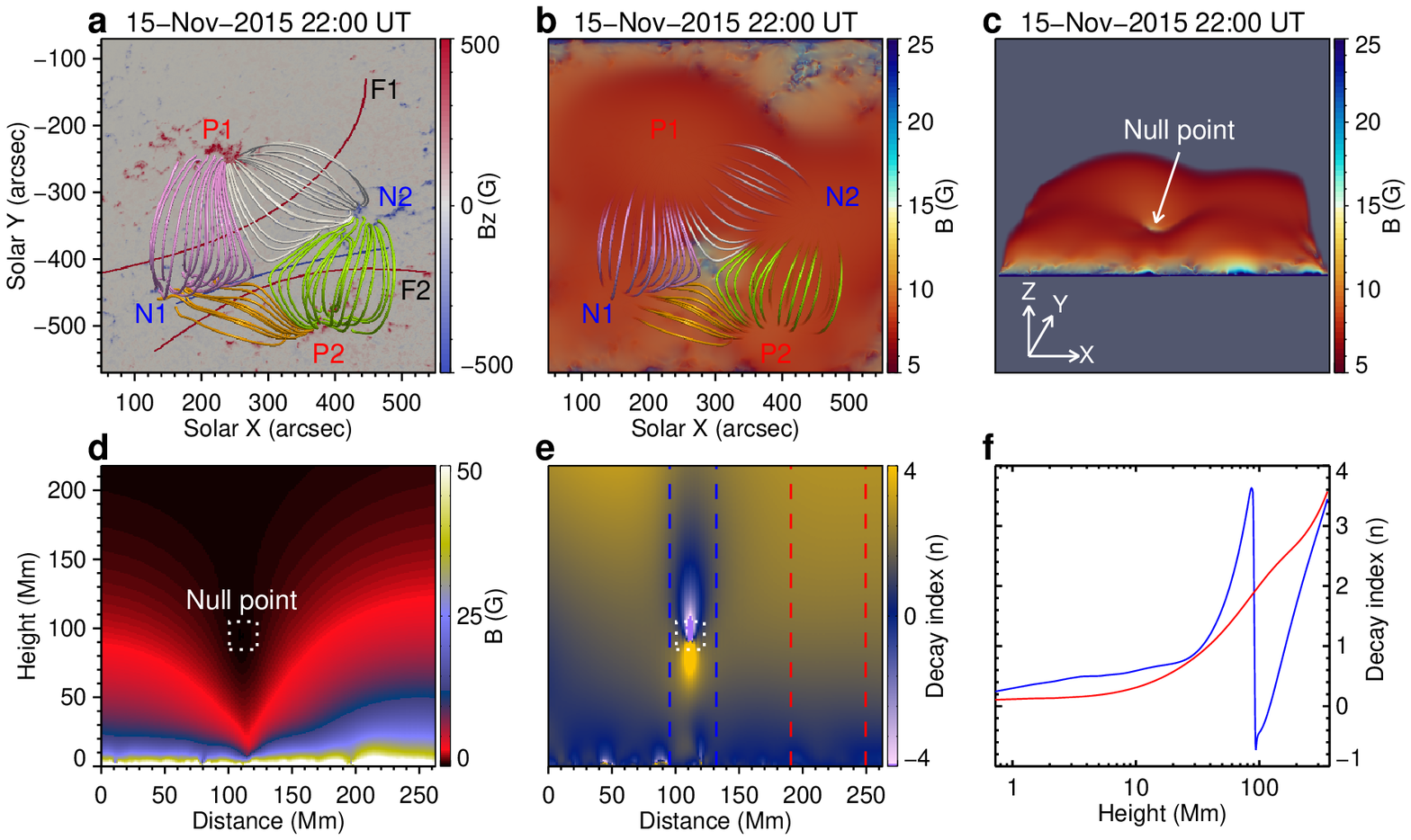}
\caption{3D magnetic fields above F1 and F2 revealed by NLFFF extrapolation at 22:00 UT on November 15.
(a): Overlying loops of F1 and F2, forming a quadrupole magnetic field configuration.
(b)--(c): Top view and side view of 3D distribution of magnetic field strength with values between 5 and 25 Gauss
in the selected cube.
(d)--(e): Distributions of magnetic field strength and decay index in the vertical plane based on the dark blue
cut denoted in (a). The null point of the quadrupole magnetic configuration is distinctly depicted at a height of
$\sim$90 Mm.
(f): The height profiles of decay index averaged within two regions defined by two blue lines and two red lines
in the vertical plane of (e), respectively.
}
\label{fig4}
\end{figure*}

\begin{figure*}
\centering
\includegraphics [width=0.99\textwidth]{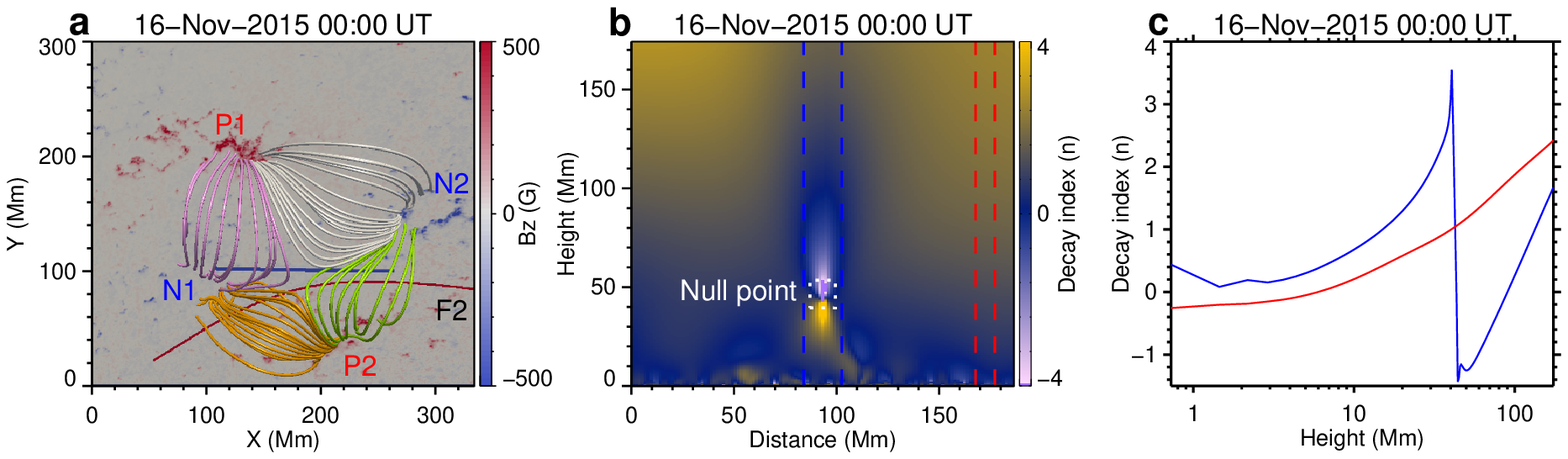}
\caption{The NLFFF extrapolation results derived from photospheric vector field of 00:00 UT on November 16.
}
\label{fig5}
\end{figure*}

Bright ribbons and coronal dimmings detected during solar eruptions essentially reflect the energy release and magnetic
topology reconfiguration resulting from successive reconnection in the corona. The two ribbons showing apparent
separation motions caused by the erupting F1 in this work can be well explained by the reconnection beneath rising F1
in the standard flare model (Priest \& Forbes 2002). However, the inward-spreading brightening IBF2 and subsequent
dimming on the south side of pre-eruption F2, as well as the synchronicity of transient brightenings appearing on both
sides of F2 ought to be attributed to a different physical mechanism: the external reconnection between the overlying
fields above erupting F1 and stable F2 (T{\"o}r{\"o}k et al. 2011). The concept of external reconnection was also
applied in previous works to explain the EUV brightenings and coronal dimming nearby a pre-eruption filament (Jiang et
al. 2011; Joshi et al. 2016; Li et al. 2017a; Wang et al. 2018). In the present work, the evolutions of inward-spreading 
brightening and its following coronal dimming beside the pre-eruption filament are unambiguously detected. They are very 
similar to the widespread secondary dimmings behind the diffuse bright fronts caused by the driven reconnections between 
the outer edge of an expanding CME magnetic filed and quiet-Sun magnetic loops (Attrill et al. 2007; Mandrini et al. 2007). 
The brightenings on both sides of stable F2 could correspond to the footpoints of newly-formed loops through the external 
reconnection between the magnetic fields above erupting F1 and stable F2. As for the dimming region behind the 
inward-spreading brightenings, it formed after the eruption of F1 but was located on the south side of pre-eruption F2. 
Compared with the core dimming region around the north footpoints of erupting F1 (not shown in the present work), this 
dimming region underwent a relatively slow density reduction of about $30\%$, which recovered soon after several hours. 
As a result, we speculate that this dimming region could be a secondary dimming and host footpoints of coronal arcades 
overlying F2, which reconnected with the expanding fields above erupting F1. This secondary dimming was produced by the 
reconnection-related magnetic fields reconfiguration, subsequent expansion or evacuation of plasma frozen in the coronal 
fields, and resultant emission reduction. Additionally, it is also possible that this secondary dimming hosted footpoints 
of large-scale loops above F1 and F2, which thus were stretched by the erupting F1. For more detailed information about 
this process, analysis of the magnetic topology above the two filaments is needed.

\subsection{Quadrupolar Magnetic Topology with a Null Point above F1 and F2}
\begin{figure*}
\centering
\includegraphics [width=0.8\textwidth]{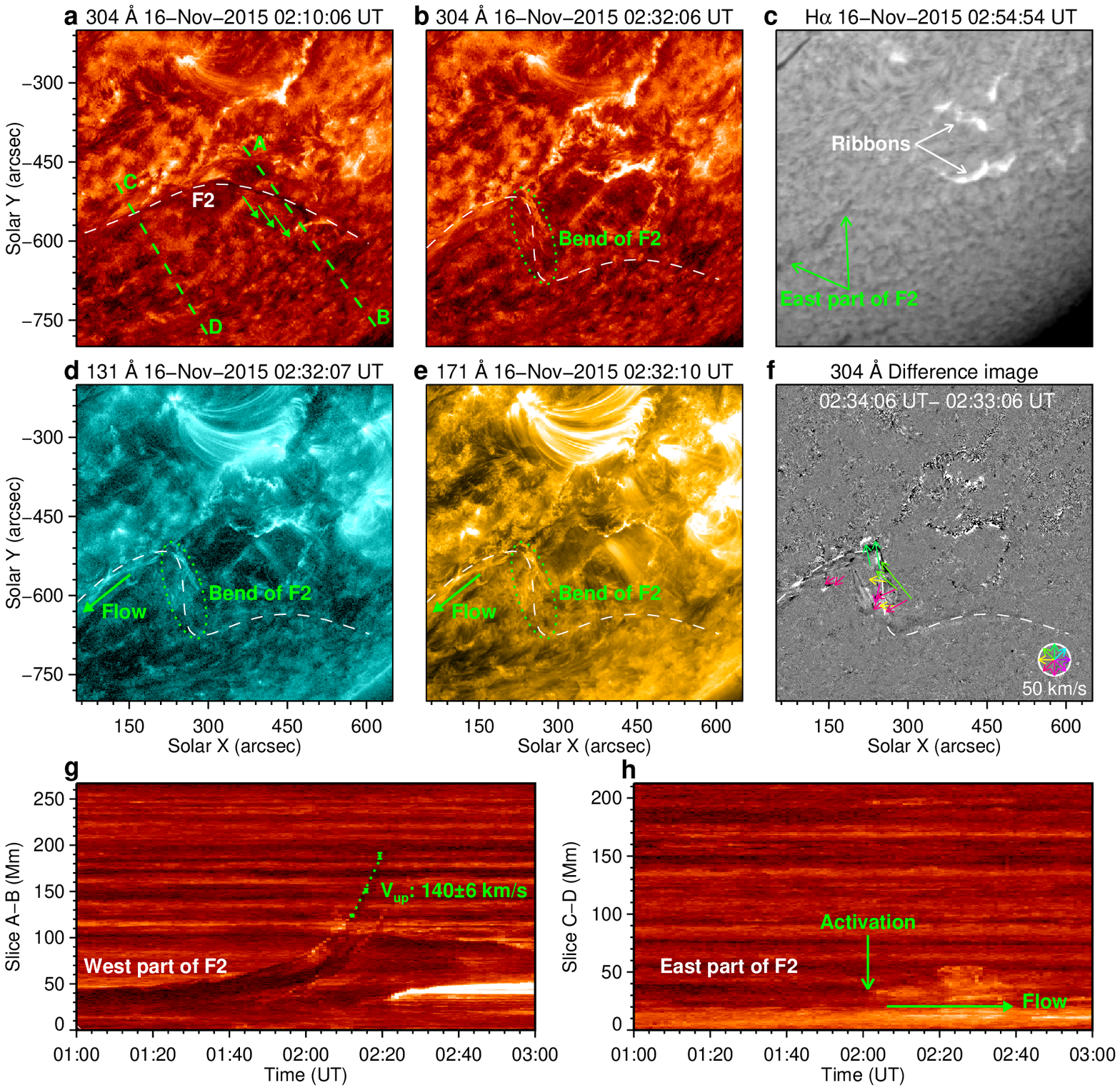}
\caption{Partial eruption of F2.
(a)--(c): 304 {\AA} and H$\alpha$ images showing partial eruption of F2. Green arrows in (a) denote eruptive
direction of the west part of F2. Green ellipse in (b) marks a bending structure in the middle part of F2 during
its rapid rising phase. White curves in (a)--(b) delineate the axis of eruptive F2.
(d)--(f): 131 {\AA}, 171 {\AA} images, and difference image of 304 {\AA} corresponding to (b). Green arrows
in (d) and (e) represent southeastward motion of the activated material of the east part of F2. The colored
arrows in (f) represent the horizontal velocity fields of filament material with values larger than 30 km s$^{-1}$
derived from 304 {\AA} images through LCT method.
(g)--(h): Time-distance plots derived from the 304 {\AA} images along lines ``A--B'' and ``C--D'' shown in (a).
The dotted oblique line in (g) delineates linear fitting of the front of eruptive F2 in its rapid rising phase.
An associated animation (figure6.mov) of 304 {\AA}, 171 {\AA}, and 131 {\AA} images, covering from 23:50 UT on
November 15 to 03:59 UT on November 16, is available online.
}
\label{fig6}
\end{figure*}

To understand the magnetic topology that leads to the features shown in Figs. 2 and 3, we extrapolate 3D magnetic fields
above F1 and F2 through NLFFF modeling based on the photospheric vector field of 22:00 UT on November 15, before the
eruption of F1. Figure 4(a) reveals a quadrupolar system above F1 and F2, consisting of four magnetic connectivities
(P1/N1 and P1/N2 above F1, P2/N2 and P2/N1 above F2). A large gradient in field line mapping is prominent in this system,
e.g., field lines emanating from P1 are at first parallel but then diverge significantly near their apexes, forming two
sets of loops that are almost antiparallel with each other and connected to N1 (pink) and N2 (white), respectively.
Field lines emanating from P2 have the same topology, forming two groups of structures (orange P2/N1 and green P2/N2).
Such magnetic topology strongly indicates the existence of a null point, which is further supported by the 3D distribution
of magnetic field strength (panels (b)--(c)). Checking distribution of field strength in the vertical plane based on the
dark blue curve marked in panel (a), we indeed find a possible null point situated at $\sim$90 Mm height (panel (d)).
Corresponding decay index map (panel (e)) and the height profiles of decay index (panel (f)) reveal that in a limited
altitude range near this null point, the decay index increases drastically, then impulsively decreases to \textbf{a}
negative value, followed by a gradual positive recovery. These features are well consistent with the existence of a null
point, where field strength becomes zero (T{\"o}r{\"o}k \& Kliem 2007).

Furthermore, we perform NLFFF extrapolation based on the photospheric vector field of 00:00 UT on November 16, when the
ribbon SRF1 produced by erupting F1 approached stable F2, and inward-spreading brightening and coronal dimming appeared
on the south side of pre-eruption F2. The extrapolation results shown in Fig. 5 reveal that at that time, the quadrupolar
system still existed there, but the null point moved southwestward in the POS. Furthermore, the height of the null point
had decreased from $\sim$90 Mm to $\sim$40 Mm.

\subsection{Physical Scenario of the Sympathetic Eruption of F1 and F2: External Reconnection and Subsequent
Reconfiguration of the Overlying Magnetic Fields}
The quadrupolar configuration with a possible null point above F1 and F2 revealed by the field extrapolation is favorable
for magnetic reconnection (Aschwanden et al. 1999; Sun et al. 2012). We infer that this configuration facilitates the
sympathetic eruptions of F1 and F2 as follows. As reported by Song et al. (2020), the eruption of F1 was triggered by
the flux cancellation within F1 channel. Then reconnection occurred beneath the erupting F1 and produced two flare ribbons
as shown in Figs. 1 and 2, as well as the associated animations. The rising F1 kept pushing its overlying magnetic fields
upwards (with a POS component of southwest), approaching the overlying fields of stable F2 under the quadrupolar
configuration shown in Fig. 4(a). Then the external magnetic reconnection occurred between pink P1/N1 above F1 and green
P2/N2 above F2 near the null point, forming new loops connecting P2 and N1. Then brightenings on both sides of stable F2
appeared around the footpoints of these newly-formed loops and showed good synchronicity (Fig. 2(d)). As the external
reconnection went on, the fields above F2 would be involved gradually from the outside to the inside (lower and lower in
height), manifesting as the inward-spreading brightening in lower atmosphere shown in Fig. 2 (corresponding to footpoint
motion of the field lines formed by the external reconnection). Besides, the external reconnection would also reconfigure
the overlying fields of F2 as follows: transporting magnetic flux from P2/N2 (green curves in Fig. 4(a)) to P2/N1 (orange
curves), weakening confinement above the west part of F2, and reinforcing confining fields above the east part of F2.
Since the magnetic flux in system P2/N2 was gradually reduced in a fixed sized space, in the term of magnetic flux
density, this process was equivalent to the expansion of magnetic system P2/N2, which usually produced the secondary
coronal dimming (Mandrini et al. 2007; Dissauer et al. 2018b). As a result, the plasma frozen in magnetic loops P2/N2
around the positive footpoints, which were within the dimming region (see Figs. 3(a3) and 4(a)), would be gradually
evacuated outward from this region. Then, the plasma density of the region above the positive footpoints of
P2/N2 kept decreasing, and a secondary coronal dimming was gradually extending there, behind the inward-spreading
brightening (Figs. 2 and 3).

Consistent well with the physical scenario proposed above, in the quadrupolar system revealed by the NLFFF extrapolation
at 00:00 UT on November 16, the null point moved southwestward in the POS and had a lower height than about 2 hours ago
(comparing Fig. 4 with Fig. 5). It indicates that the field above erupting F1 (pink curves in Figs. 4 and 5) kept
approaching the overlying fields of F2 (green curves), leading to successive external reconnection around the null point.
As the reconnection continued, the outermost layer of the fields above F1 and F2 would be removed gradually, and the
height of the reconnection site as well as the null point kept decreasing.

Moreover, the reconfiguration of overlying fields of F2 is further supported by the observations that F2 eventually
underwent a partial eruption (Fig. 6 and the associated animation). Suffering impacts from the erupting F1, the west
part of F2 started to rise slowly from 00:30 UT on November 16. Around 02:00 UT, the west part of F2 went through a
robust acceleration and successfully erupted southwestward in the POS (Fig. 6(a)), producing two flare ribbons (panels
(b) and (c)). However, at the east part of F2, the filament material was activated but confined within the filament 
channel and then moved southeastward (panels (c)-(e)). It is notable that during the rapid rising phase of F2, a 
bending structure appeared in its middle part (panels (b), (d), and (e)). The difference image of 304 {\AA} channel
clearly depicts the bending axis of F2 (white curve in panel (f)). Moreover, the velocity fields derived from LCT
method show that filament materiel in the bending region basically has two apparent motion patterns: northward one
and southward one. Here we suggest that as a result of reconnection-related reconfiguration of the overlying fields
aforementioned, the eruption of the east part of F2 was blocked by its enhanced overlying fields, thus forming
a bending structure between itself and the successfully erupting west part, where the overlying fields have been
significantly weakened. The material within the bending region falls to the east part of F2 and then flows along the
remanent filament fields. The time-distance plots in Figs. 6(g)--6(h) show that the west part of F2 eventually erupted
with a POS velocity of 140 $\pm$ 6 km s$^{-1}$ while the material in the east part moved along the filament axis after
being activated.

Noting that owing to the temporal overlap with the eruption of F1 and projection effect, along with that the external
reconnection is expected to be relatively weak, observational signatures during the external reconnection process,
such as the quasi-periodic pulsation (QPP) around the reconnection moment (Zhou et al. 2019; Chen et al. 2019) or
emission enhancements in EUV and X-ray channels (Yan et al. 2018; Zou et al. 2020) near the possible null point, are
difficult to be unambiguously distinguished from dominant features caused by the robust reconnection beneath erupting
F1 in this event. Moreover, the loops corresponding to the overlying fields P1/N1, P2/N2, and newly-formed P2/N1 were
also not distinct in the AIA imaging observations. The absence of these loops could be caused by the weakness of the
overlying fields and the external reconnection occurring between them. Although we have detected and analyzed some
pieces of important evidence, which could constitute an integrated chain to support occurrence of the external 
reconnection, the scenario proposed in the present work ought to be regarded as one alternative rather than exclusive 
explanation due to the absence of the observational signatures mentioned above. For example, the secondary coronal 
dimming reported here could also host footpoints of large-scale loops above F1 and F2, which were stretched by the 
erupting F1 and caused the formation of the secondary dimming.

\section{Summary}
We summarize our results as follows:
\begin{enumerate}
\item On 2015 November 15-16, two filaments (north F1 and south F2) separated by a narrow region of negative polarity
successively erupted, showing distinct sympathetic characteristics. The eruption of each filament was associated with a
B-class flare and a CME.

\item When the southward-spreading ribbon (SRF1) produced by first-erupting F1 approached stable F2, a relatively weak
brightening (IBF2) was observed on the south side of F2 and then kept spreading northward, inward approaching F2.
Moreover, brightenings could be detected on both sides of F2 and showed good synchronicity.

\item Behind the inward-spreading brightening, a distinct secondary coronal dimming region was extending,
where the mean value of plasma density decreased by about $30\%$ and that of temperature declined by $10\%$.

\item NLFFF extrapolations reveal that fields above pre-eruption F1 and F2 constituted a quadrupolar magnetic system
with a possible null point. Moreover, the null point kept moving towards F2 and descending within the following hours.

\item F2 eventually partially erupted and formed a bending structure between the erupting west part and the
remained east part, which indicated non-uniform distribution of the constraint fields above F2.

\item We propose that the rising F1 pushed its overlying fields towards the fields above stable F2 and caused successive
external reconnection near the null point, resulting in simultaneous brightenings on both sides of stable F2 and the
motion of the null point. From outside to inside (lower and lower in height), the fields above pre-eruption F2 were
gradually involved in the reconnection, manifesting as the inward-spreading brightening and dimming on the south side
of F2. The external reconnection could then reconfigure the confining fields above F2, resulting in its subsequent
partial eruption.
\end{enumerate}

The event analyzed here presents abundant sympathetic characteristics. Focusing on important observational evidence
(e.g., inward-spreading brightening and secondary dimming region on the south side of the pre-eruption F2, the subsequent
partial eruption of F2, and the bending structure formed between the erupting west part and the remained east part of F2),
we aim to reveal specific physical process of the interaction between erupting F1 and pre-eruption F2 and the subsequent
eruption of F2. Based on more thorough NLFFF extrapolations with a time sequence, a quadrupolar magnetic system with a
possible null point above the two filaments and its evolution are revealed. Here we report evolutions of the inward-spreading 
brightening and secondary coronal dimming beside the pre-eruption filament, reconfiguration of the quadrupolar magnetic 
system with a possible null point, a shift of the null-point to a lower altitude, and the resultant partial eruption of 
F2 during the sympathetic eruptions of two filaments. From the aspects of pre-existing magnetic topology in favor of 
reconnection, observational evidence of the external reconnection, and the reconnection-related reconfiguration of overlying 
fields, we propose an integrated evidence chain to unambiguously demonstrate the key roles of external magnetic reconnection 
between overlying fields and the resultant reconfiguration of magnetic topology on the sympathetic eruptions of two filaments. 
The characteristics of emission and magnetic topology investigated in the present work should be taken into account when 
studying the sympathetic eruptions of two filaments in future works.

\begin{acknowledgements}
The authors are grateful to the anonymous referee for the constructive suggestions improving this paper.
Y.H. appreciates Dr. Shuhong Yang, Dr. Xiaoshuai Zhu, and Dr. Shuo Yang for valuable discussions.
The data are used courtesy of \emph{SDO}, GONG, \emph{SOHO}, and \emph{GOES} science teams. \emph{SDO} is a mission of
NASA's Living With a Star Program. This work is supported by the Strategic Priority Research Program of the Chinese
Academy of Sciences (XDB41000000), the National Natural Science Foundation of China (11903050, 11790304, 11773039, 11533008,
11873059, 11673035, 11673034, and 11790300), the National Key R\&D Program of China (2019YFA0405000), the NAOC Nebula
Talents Program, the Youth Innovation Promotion Association of CAS (2017078), Young Elite Scientists Sponsorship Program
by CAST (2018QNRC001), and Key Programs of the Chinese Academy of Sciences (QYZDJ-SSW-SLH050). Z.S. acknowledges support
from the open topic of the Key Laboratory of Solar Activities of the Chinese Academy of Sciences (KLSA201902).
\end{acknowledgements}

%
%

\clearpage

\end{document}